\documentclass[pra,aps,tightenlines,byrevtex,twocolumn,showpacs]{revtex4} 

\usepackage{epsfig}

\newcommand {\bra}[2] {\mbox{}_{#2}\langle #1 |} 
\newcommand {\ket}[2] {| #1 \rangle_{#2}} 
\newcommand {\braket}[4] {\mbox{}_{#3}\langle #1 | #2 \rangle_{#4}} 
 
\newcommand {\half} {\frac{1}{2}} 
\newcommand {\fig}[1] {Fig.~\ref{#1}} 
\newcommand {\figwidth} {90mm} 
\newcommand {\Ref}[1] {Ref.~\cite{#1}}

\begin{document}

\title{Teleportation with the entangled states of a beam splitter}

\author{P. T. Cochrane} \email{cochrane@physics.uq.edu.au}

\author{G. J. Milburn}

\affiliation{Department of Physics, The University of Queensland, 
St.~Lucia, Queensland 4072, Australia} \date{\today}

\begin{abstract}
We present a teleportation protocol based upon the entanglement 
produced from Fock states incident onto a beam splitter of arbitrary 
transmissivity.  The teleportation fidelity is analysed, its trends 
being explained from consideration of a beam splitter's input/output 
characteristics.
\end{abstract}


\maketitle

Entanglement is a resource with which to perform quantum information 
processing tasks, such as quantum 
computing~\cite{DiVincenzo:1995:1,Grover:1997:1,Ekert:1998:1,Jozsa:1998:1}, 
quantum error correction~\cite{Bennett:1996:1,Knill:1997:1}, dense 
coding~\cite{Bennett:1992:2,Braunstein:2000:3} and quantum 
teleportation~\cite{Bennett:1993:1,Furusawa:1998:1,Bouwmeester:1997:1,%
Braunstein:1998:2}.  In particular, teleportation has generated a lot 
of interest since it was first proposed~\cite{Bennett:1993:1} and 
demonstrated~\cite{Furusawa:1998:1,Bouwmeester:1997:1}.  There are 
many protocols for teleportation using both discrete and continuous 
variables~\cite{Braunstein:1998:2,Ralph:1998:1,Bennett:1993:1,%
Cochrane:2000:2}, nevertheless all are based upon the original 
proposal.  For further related work the reader is directed to 
references~\cite{Ralph:1999:2,Milburn:1999:1,Clausen:2000:2,%
Braunstein:2000:2,DelRe:2000:1,Enk:1999:2,Trump:2001:1,Yu:2000:2}.

In this paper we generalise and expand upon results of previous 
work~\cite{Cochrane:2000:2}, showing how harmonic oscillator states 
entangled on a beam splitter may be used as an entanglement resource 
for teleportation.  We describe the teleportation protocol and derive 
the fidelity of output showing its behaviour as a function of the 
difference in photon number incident to the beam splitter and the 
transmission properties of the beam splitter.  The average fidelity 
trends are as expected from a simple consideration of the beam 
splitter.

The process of teleportation can be explained in general terms as 
follows: There are two parties who wish to communicate quantum 
information between one another; a sender, Alice, and a receiver, Bob.  
Alice and Bob initially share one part each of a bipartite entangled 
system.  Alice also has a particle of an unknown quantum state, this 
being the information she wishes to send to Bob.  She sends this 
information by making \emph{joint} measurements on her part of the 
entangled pair and the unknown particle, and then sending the results 
of these measurements to Bob via the classical channel.  Bob can then 
recreate the unknown quantum state perfectly (in principle) after 
performing local unitary transformations on his part of the entangled 
pair.  The important point is that, in principle, perfect transmission 
of quantum information is possible between spatially separated points 
but only with the help of quantum entanglement.

There are many processes involved in performing teleportation; the 
measurements made by Alice, the transmission of the classical 
information, and the transformations made by Bob.  If one assumes that 
these processes are all performed perfectly, then the only influence 
on the efficacy of teleportation will be the quality of the 
entanglement.

\begin{figure}
\centerline{\epsfig{file=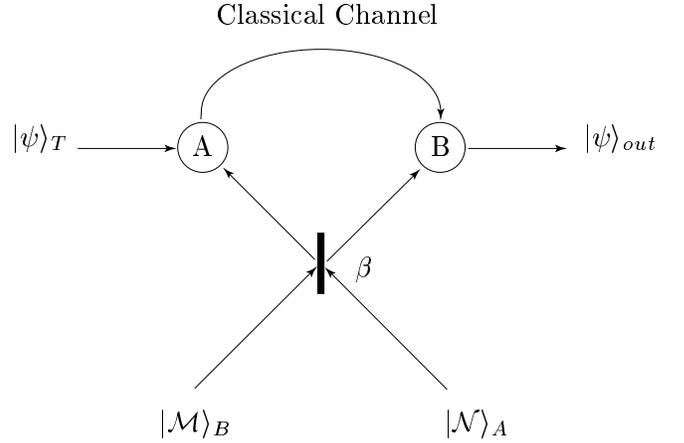, width=\figwidth}}
\caption{Schematic experimental setup for teleportation protocol.  
$\ket{\mathcal{M}}{B}$ and $\ket{\mathcal{N}}{A}$ are input Fock 
states to a beam splitter of transmissivity $\beta$.  The sender of 
the target state $\ket{\psi}{T}$ is at A and the receiver is at B. The 
state exiting the teleportation process is denoted by 
$\ket{\psi}{out}$.}
\label{fig:expt}
\end{figure}

Consider the experiment shown schematically in \fig{fig:expt}.
Two Fock states, number $\mathcal{N}$ in mode $A$, and $\mathcal{M}$ 
in mode $B$, are incident on a beam splitter with transmissivity 
described by the parameter $\beta$.  Mode $A$ goes to Alice and mode 
$B$ goes to Bob.  Alice makes joint number sum and phase difference 
measurements~\cite{Cochrane:2000:2} on the target and mode $A$.  She 
sends the results of these measurements to Bob via the classical 
channel, who then applies the relevant unitary transformations on his 
mode to attempt to recreate the target state at his location.

The input Fock states are entangled via the beam splitter interaction; 
described by
\begin{equation}
\ket{\psi}{AB} = e^{i\beta (a^\dagger b + b^\dagger a)/2} 
\ket{\mathcal{N}}{A} \ket{\mathcal{M}}{B},
\end{equation}
where $a$, $a^\dagger$, $b$ and $b^\dagger$ are the usual boson 
annihilation and creation operators for modes $A$ and $B$ 
respectively.  The variable $\beta$ describes the transmissivity of 
the beam splitter; $\beta=0$ corresponds to all transmission and no 
reflection, $\beta=\pi$ corresponds to all reflection and no 
transmission and $\beta=\pi/2$ corresponds to a 50:50 beam splitter.  
When the total photon number is fixed, these states can be written in 
a pseudo-angular momentum algebra, allowing the resource to be 
expanded in terms of eigenstates of constant number sum.  The resource 
state is:
\begin{equation}
\rho_{AB} = \sum_{n,n' = 0}^{2N} d_{n-N} d_{n'-N}^* 
\ket{n}{A}\bra{n'}{} \otimes \ket{2N - n}{B}\bra{2N - n'}{}.
\label{eq:res}
\end{equation}
The $d_{n-N}$ are
\begin{equation}
d_{n - N} = e^{-i\frac{\pi}{2}(n-N - m)} D_{n-N,m}^{2N}(\beta),
\end{equation}
where $m = (\mathcal{N}-\mathcal{M})/2$ is the incident photon number 
difference and the $D_{m',m}^j(\beta)$ being the rotation matrix 
coefficients~\cite{Biedenharn:1981:1} given by
\begin{eqnarray}
D_{m',m}^j(\beta) = 
\left[(j+m')!(j-m')!(j+m)!(j-m)!\right]^\half\nonumber\\
\times \sum_s \frac{(-1)^{m'-m+s} 
\left(\cos\frac{\beta}{2}\right)^{2j+m-m'-2s} 
\left(\sin\frac{\beta}{2}\right)^{m'-m+2s}}{(j+m-s)!s!(m'-m+s)!(j-m'-s)!}.
\end{eqnarray}
The variable $s$ ranges over all possible values such that the 
factorials are positive.  The resource states are eigenstates of 
number-sum and tend to eigenstates of phase-difference in the limit of 
large total photon number (for details see \Ref{Cochrane:2000:2}).

The quality of information transfer is measured by the overlap between 
the target state and the output state.  This is the \emph{fidelity} 
which we define by
\begin{equation}
F = \bra{\psi}{T} \rho_{out,B} \ket{\psi}{T}.
\end{equation}

We now show the mechanics of our teleportation protocol in order to 
calculate the teleportation fidelity.  We teleport an arbitrary state 
of the form
\begin{equation}
\rho_T = \sum_{m,m' = 0}^\infty c_m c_{m'}^* \ket{m}{T}\bra{m'}{}.
\end{equation}
The subscript $T$ emphasises that this is the ``target'' state and the 
$c_m$ are chosen such that the state is normalised.  The total state 
of the system is the tensor product between this and $\rho_{AB}$.
\begin{equation}
\rho_{TAB} = \rho_T \otimes \rho_{AB}
\end{equation}
Alice makes joint measurements of number-sum (yielding result $q$) and 
phase-difference (result $\phi_-$) on the target and her half of the 
entangled pair, mode $A$.  The state of the system conditioned on 
these measurements is
\begin{eqnarray}
\rho^{(q,\phi_-)} &=& \sum_{w,y,x',z'=0}^\infty e^{i(y-w+x'-z')\phi_-} 
\delta_{w-q+y} \delta_{q-x'-z'} \\ \nonumber &&\times 
\ket{w}{T}\bra{x'}{}\otimes\ket{y}{A}\bra{z'}{}\\ \nonumber &\otimes& 
\frac{1}{P(q)} \sum_{n,n'=0}^{min(q,2N)} e^{-2i(n-n')\phi_-} c_{q-n} 
c_{q-n'}^* d_{n-N} d_{n'-N}^* \\ \nonumber &&\times \ket{2N-n}{B} 
\bra{2N-n'}{},
\end{eqnarray}
where
\begin{equation}
P(q) = \sum_{n = 0}^{min(q,2N)} |c_{q-n}|^2 |d_{n-N}|^2,
\end{equation}
is the probability of measuring a given number-sum result $q$.  Alice 
transmits the results of these measurements to Bob via the classical 
channel.  Bob makes the amplification operations
\begin{eqnarray}
\ket{2N - n}{B} \rightarrow \ket{q - n}{B}\\ \nonumber \bra{2N - 
n'}{B} \rightarrow \bra{q - n'}{B}
\end{eqnarray}
and the phase shift $e^{2i(n - n')\phi_-}$.  The unitary amplification 
operation is described in~\cite{Wiseman:1995:1} and in more detail 
in~\cite{Wiseman:1994:1}; other amplification techniques are discussed 
by Yuen~\cite{Yuen:1986:1} and Bj\"ork and 
Yamamoto~\cite{Bjork:1988:1}.  The amplifications and phase 
transformations complete the protocol.  Bob's state is then
\begin{eqnarray}
\rho_{out,B} &=& \frac{1}{P(q)} \sum_{n,n' = 0}^{min(q,2N)} c_{q-n} 
c_{q-n'}^* d_{n-N} d_{n'-N}^* \\ \nonumber &&\times\ket{q - 
n}{B}\bra{q - n'}{},
\end{eqnarray}
and the teleportation fidelity is,
\begin{equation}
F(q) = \frac{1}{P(q)} \sum_{n,n'=0}^{min(q,2N)} |c_{q-n}|^2 
|c_{q-n'}|^2 d_{n-N} d_{n'-N}^*.
\end{equation}

Note that the fidelity is dependent upon the number sum measurement 
result ($q$).  To obtain an overall figure of merit for the protocol 
we remove this dependence by defining the \emph{average fidelity},
\begin{equation}
\bar{F} = \sum_q P(q) F(q).
\end{equation}
For our protocol this is,
\begin{equation}
\bar{F} = \sum_{q = 0}^\infty \sum_{n,n'=0}^{min(q,2N)} |c_{q-n}|^2 
|c_{q-n'}|^2 d_{n-N} d_{n'-N}^*.
\end{equation}
If one sets $\mathcal{N} = \mathcal{M}$, this result is identical to 
that obtained in \Ref{Cochrane:2000:2} without decoherence.

\begin{figure}
\centerline{\epsfig{file=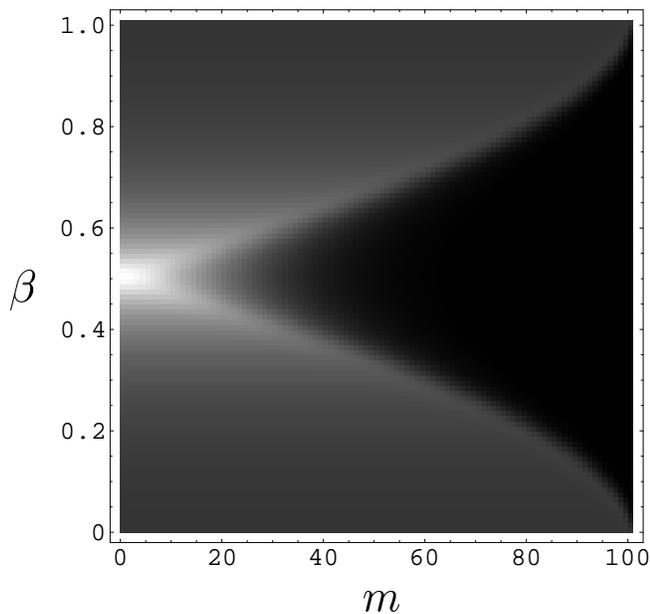, width=\figwidth}}
\caption{Density plot of average fidelity as a function of $m$ and 
$\beta$ for an ``even cat'' target state of amplitude $\alpha = 3$ and 
total photon number of 100.  $\beta$ is in units of $\pi$; $\beta = 
\pi/2$ corresponds to a 50:50 beam splitter; black corresponds to 
zero and white to unity.}
\label{fig:Fbsbetam}
\end{figure}

Teleportation fidelity for transmission of an ``even cat'' target 
state of amplitude $\alpha = 3$ is shown in \fig{fig:Fbsbetam}.  An 
even cat state is the even superposition of two coherent states of 
equal amplitude but opposite phase~\cite{Cochrane:1999:1}, i.e.
\begin{equation}
\ket{cat}{even} = \frac{\ket{\alpha}{} + \ket{-\alpha}{}}{\sqrt{2 + 
2\exp(-2|\alpha|^2)}}.
\label{eq:}
\end{equation}
Many of the trends shown in \fig{fig:Fbsbetam} can be explained by 
simple consideration of a beam splitter.  As the beam splitter becomes 
more biased ($\beta$ tends to either 0 or $\pi$), the outgoing photons 
are partitioned less evenly and the entanglement resource is 
distorted.  This is evident by the average fidelity decreasing to the 
classical level\footnote{The classical level is defined as the 
fidelity achieved with this protocol using no entanglement resource.} 
at $\beta = 0$ and $\beta = \pi$.  At these extremes the setup is 
completely biased with all incident photons being sent in one 
direction, so there are no phase correlations between the modes above 
the classical level.  Changing the photon number difference also 
changes the partitioning of outgoing photons, hence the fidelity 
decreases with increasing $m$ for the same reasons outlined above.  
The input photon number difference and beam splitter transmissivity 
can have opposing photon partitioning effects, thereby keeping the 
fidelity high.  This is evident by the ``ridges'' of the fidelity 
surface.  The ridges decrease in height with increasing $m$ implying 
that although the two biases are in opposition, the resource is still 
being distorted.  

\begin{figure}
\centerline{\epsfig{file=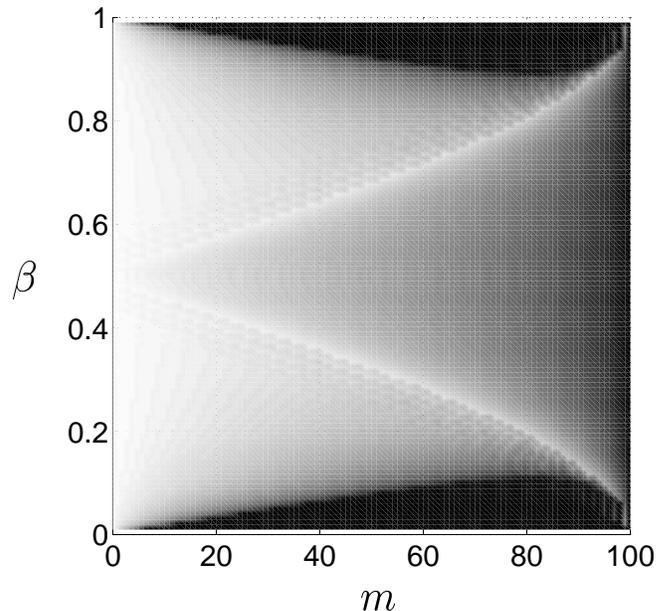, width=\figwidth}}
\caption{Density plot of phase difference $\phi_-$ corresponding to a 
maximum joint phase probability $P_{max}(\phi_-,\beta,m)$, as a 
function of beam splitter transmissivity $\beta$ and input photon 
number difference $m$.  The ridge structure here helps to explain the 
ridge structure of the average fidelity $\bar{F}$ as a function of the 
same variables.  Black corresponds to zero, white to $\pi/2$.}
\label{fig:phiAtMax}
\end{figure}

We can show why the ridges occur in a more quantitative fashion with 
the aid of the joint phase probability of the resource state.  This is
\begin{equation}
P(\phi_1,\phi_2,\beta,m) =
\left|\bra{\phi_1}{}\braket{\phi_2}{\psi}{}{AB}\right|^2
\label{eq:jointPhaseProb}
\end{equation}
where the $\ket{\phi_j}{}$ are the phase states
\begin{equation}
\ket{\phi}{j} = \sum_{n = 0}^\infty e^{-i\phi_j n} \ket{n}{}.
\end{equation}
Equation~(\ref{eq:jointPhaseProb}) can be written in a more explicit 
form;
\begin{equation}
P(\phi_-,\beta,m) = \left| \sum_{n=0}^{2N} e^{in\phi_-} 
d_{n-N}(\beta,m) \right|^2,
\end{equation}
where $\phi_-$ is the phase difference $\phi_1 - \phi_2$.  This is a 
function of three variables: phase difference $\phi_-$, beam splitter 
transmissivity $\beta$ and photon number difference $m$, and is 
consequently not easy to analyse graphically.  However, if one finds 
the maximum of $P(\phi_-,\beta,m)$ over $\phi_-$ for given $\beta$ and 
$m$ (we call this quantity $P_{max}$), and the value of $\phi_-$ that 
corresponds to this maximum, then we obtain more easily 
interpretable information.  We show in \fig{fig:phiAtMax} the value of 
$\phi_-$ corresponding to $P_{max}$ as a function of 
$\beta$ and $m$.
This function shows the same ridge structure as \fig{fig:Fbsbetam}.  
When $\beta = \pi/2$ and $m = 0$, the average fidelity is a maximum 
and the joint phase probability density has a maximum at $\phi_- = 
\pi/2$.  For other values of $\beta$ and $m$ the ridges in the average 
fidelity correspond to where the phase distribution has a maximum 
near $\pi/2$ and where the protocol is therefore better.

Testing our results experimentally will be difficult in the optical 
regime.  However, recent 
experiments~\cite{Varcoe:2000:1,Benson:2000:1,Kurtsiefer:2000:1} 
showing generation of Fock states, and proposals using alternative 
technologies~\cite{Foden:2000:2,Brunel:1999:1} indicate some future 
possibility of exploring the ideas presented here.

We have shown how Fock states entangled on a beam splitter may be used 
as an entanglement resource for teleportation in the case of arbitrary 
beam splitter properties and arbitrary input Fock states.  We have 
studied how varying the beam splitter transmissivity and input photon 
number difference influences the average fidelity.  The results are 
consistent with an analysis of how entanglement varies with these 
parameters.

\begin{acknowledgments}
PTC acknowledges the financial support of the Centre for Laser Science 
and the University of Queensland Postgraduate Research Scholarship.  
PTC also thanks W.~J. Munro for helpful discussions.
\end{acknowledgments}

\end{document}